STATISTICAL, NONLINEAR,
AND SOFT MATTER PHYSICS

# High Temperature Equation of State of Metallic Hydrogen

V. T. Shvets

*Odessa State Academy of Refrigeration, Odessa, 65026 Ukraine*
**e-mail: valtar@paco.net*
Received May 5, 2006

**Abstract**—The equation of state of liquid metallic hydrogen is solved numerically. Investigations are carried out at temperatures from 3000 to 20000 K and densities from 0.2 to 3 mol/cm$^3$, which correspond both to the experimental conditions under which metallic hydrogen is produced on earth and the conditions in the cores of giant planets of the solar system such as Jupiter and Saturn. It is assumed that hydrogen is in an atomic state and all its electrons are collectivized. Perturbation theory in the electron–proton interaction is applied to determine the thermodynamic potentials of metallic hydrogen. The electron subsystem is considered in the random-phase approximation with regard to the exchange interaction and the correlation of electrons in the local-field approximation. The proton–proton interaction is taken into account in the hard-spheres approximation. The thermodynamic characteristics of metallic hydrogen are calculated with regard to the zero-, second-, and third-order perturbation theory terms. The third-order term proves to be rather essential at moderately high temperatures and densities, although it is much smaller than the second-order term. The thermodynamic potentials of metallic hydrogen are monotonically increasing functions of density and temperature. The values of pressure for the temperatures and pressures that are characteristic of the conditions under which metallic hydrogen is produced on earth coincide with the corresponding values reported by the discoverers of metallic hydrogen to a high degree of accuracy. The temperature and density ranges are found in which there exists a liquid phase of metallic hydrogen.

PACS numbers: 64.10.+h, 65.50.+m, 71.15.Nc, 71.10.+x

**DOI:** 10.1134/S1063776107040164

## 1. INTRODUCTION

The possibility of existence of metallic hydrogen was first predicted in 1935 [1]. However, its actual discovery and the detailed investigation of its electric resistivity as a function of pressure and temperature dates back to 1996 [2]. In these investigations, liquid molecular hydrogen was subject to a shock compression up to pressures of 0.93–1.80 Mbar at temperatures of 2200–4400 K. At a pressure of 1.4 Mbar and a temperature of 3000 K, a metal-to-insulator transition was observed. Actually, this was a metal–semiconductor transition, because the energy band-gap in the molecular hydrogen did not vanish but only decreased from 15 to 0.3 eV, virtually becoming equal to the sample temperature. Note that experimental and theoretical investigations of predicted metallic hydrogen were also carried out earlier. For example, in [3], the authors measured the electric resistivity of molecular hydrogen at much lower pressures of 0.1–0.2 Mbar. The resistivity exhibited exponential dependence on temperature, which is characteristic of semiconductors with energy band-gap of 12 eV. The first detailed investigation of the equation of state of metallic hydrogen in crystalline state at low temperatures was carried out in 1971 [4]. In 1978, the first report appeared on the discovery of metallic hydrogen [5] at a pressure of 2 Mbar.

Today, the equilibrium properties of metallic hydrogen are being intensively investigated [6–9]. The significance of these investigations is largely attributed to the fact that certain equilibrium characteristics of metallic hydrogen, such as density and temperature, are measured with regard to the extreme conditions of its existence under terrestrial conditions. On the other hand, such an important characteristic as pressure can be calculated. An essential feature of the investigations of the equilibrium properties of metallic hydrogen is the application of the nearly free-electron model. We also used this model for calculating the electric conductivity of metallic hydrogen [10]. In the present paper, we assume that hydrogen is in metallic state with zero energy band-gap. Such a state is realized either at high pressures or at high temperatures. Note that the core of Jupiter, whose radius is equal to half the radius of the planet itself, consists of hydrogen at pressure of 3–40 Mbar and temperature of 10000–20000 K.

## 2. HAMILTONIAN

In the nearly free-electron approximation, the Hamiltonian of the electron subsystem of metallic hydrogen can be taken in the form similar to the Hamiltonian of simple liquid metals [11]:

$$H = H_i + H_e + H_{ie}. \quad (1)$$

The Hamiltonian of the proton subsystem has the form

$$H_i = \sum_{n=1}^{N} T_n + \frac{1}{2\mathcal{V}} \sum_{\mathbf{q}} V(q)[\rho^i(\mathbf{q})\rho^i(-\mathbf{q}) - N]. \quad (2)$$

The first term on the right-hand side describes the kinetic energy of protons and the second, the energy of the Coulomb interaction of protons. Here, $\mathcal{V}$ is the volume of the system; $N$ is the amount of protons in the system; $T_n$ is the kinetic energy of the $n$th proton, $V(q)$ is the Fourier image of the Coulomb proton–proton, electron–electron, and electron–proton interactions; and $\rho^i(\mathbf{q})$ is the Fourier transform of the density of protons. For sufficiently high temperatures that are dealt with below, the proton subsystem can be assumed to be classical. Since the electron gas is strongly degenerate for all temperature values considered, it is expedient to apply the representation of secondary quantization in plane waves to describe the electron gas. In this case,

$$H_e = \sum_{\mathbf{k}} \varepsilon_k a_{\mathbf{k}}^+ a_{\mathbf{k}}$$
$$+ \frac{1}{2\mathcal{V}} \sum_{\mathbf{q}} V(q)[\rho^e(\mathbf{q})\rho^e(-\mathbf{q}) - N]. \quad (3)$$

The first term on the right-hand side describes the kinetic energy of the electron gas and the second term describes the Coulomb energy of interaction between electrons. Here, $a_{\mathbf{k}}^+$ and $a_{\mathbf{k}}$ are the operators of birth and annihilation of electrons in the state with wave vector $k$, $\varepsilon_k$ is the energy of a free electron, $m$ is its mass, $\rho^e(\mathbf{q})$ is the Fourier transform of the operator of electron density, and $N$ is the operator of the number of electrons.

The Hamiltonian of the Coulomb interaction between electrons and protons is given by

$$H_{ie} = \frac{1}{\mathcal{V}} \sum_{\mathbf{q}} V(q)\rho^i(\mathbf{q})\rho^e(-\mathbf{q}). \quad (4)$$

The condition of electrical neutrality of the system can be taken into account in the original Hamiltonian by dropping out the term with $\mathbf{q} = 0$ in each term.

## 3. INTERNAL ENERGY

The internal energy of a system can be obtained by averaging the Hamiltonian over the Gibbs canonical ensemble:

$$E = \langle H \rangle = \langle H_i \rangle + \langle H_e \rangle + \langle H_{ie} \rangle. \quad (5)$$

The contribution of the proton subsystem to the internal energy is expressed as

$$E_i = \langle H_i \rangle = N\frac{3}{2}k_B T + N\frac{1}{2\mathcal{V}} {\sum_{\mathbf{q}}}' V(q)[S^i(q) - 1]. \quad (6)$$

Here, the prime denotes that the term with $q = 0$ is missing in the sum, and $T$ is the absolute temperature of the system. The last contribution is called the Madelung energy; the accuracy to which it is calculated depends on the accuracy of the approximation used for the statistical structure factor $S^i(q)$ of the proton subsystem. As the latter factor, we will use below the structure factor of a system of hard spheres:

$$S^i(q) = [1 - nC(q)]^{-1}, \quad (7)$$

where $C(q)$ is the Fourier transform of the direct correlation function:

$$C(q) = -4\pi\sigma^3 \int_0^1 (\alpha + \beta x + \gamma x^3)\frac{\sin(q\sigma x)}{q\sigma x} x^2 dx, \quad (8)$$

$$\alpha = \frac{(2\eta+1)^2}{(1-\eta)^4}, \quad \beta = -6\eta\frac{(1+0.5\eta)^2}{(1-\eta)^4}, \quad \gamma = \frac{1}{2}\eta\alpha.$$

Here, $n$ is the density of protons, $\sigma$ is the diameter of hard spheres, and $\eta$ is the packing fraction.

It is convenient to consider the energy of the electron subsystem and the energy of the interaction between the electron and proton subsystems simultaneously. The sum of these energies, the energy of the ground state of the electron gas in the field of protons, can be expanded in powers of the electron–proton interaction [12]:

$$E_e = \langle H_e \rangle + \langle H_{ie} \rangle = \sum_{n=0}^{\infty} E_n. \quad (9)$$

In turn, in each order in the electron–proton interaction, the relevant term should be expanded in a series in the electron–electron interaction. The zero-order term in the electron–proton and electron–electron interactions is the kinetic energy of an ideal electron gas. At low temperatures ($k_B T/\varepsilon_F \ll 1$), we have

$$E_{0e} = \sum_{\mathbf{k}} \varepsilon_k n(k) = N\frac{3}{5}\varepsilon_F = N\frac{1.105}{r_s}. \quad (10)$$

Here, we introduced the Bruckner parameter $r_s$, equal to the radius of a sphere whose volume coincides with the volume of the system per electron, and $\varepsilon_F$ is the Fermi energy. In the first order in the electron–electron interaction, the contribution to energy is called the Hartree–Fock energy. To determine this energy, one should take into account the contribution of the electron–electron interaction to the Hamiltonian of the system, in which it suffices to take the structure factor of an ideal electron gas as the structure factor $S^e(q)$ [13]. As a result, we obtain

$$E_{\text{HF}} = N\frac{1}{2\mathcal{V}}\sum_{\mathbf{q}}' V(q)[S_0^{\text{e}}(q) - 1] = -N\frac{0.458}{r_{\text{s}}}. \quad (11)$$

The higher order terms in the electron–electron interaction are called correlation energy. The problem of taking into consideration these terms still remains open. A conventional approach consists in applying the Nosier–Pines interpolation formula [13, 14]

$$E_{\text{cor}} = N(-0.058 + 0.016\ln r_{\text{s}}). \quad (12)$$

Since the system is electrically neutral, the first-order term in the electron–proton interaction in the ground state energy of the electron gas in metallic hydrogen is missing. The second- and higher order terms in the electron–proton interaction, the so-called band structure energy, are expressed as

$$E_n = \frac{N}{\mathcal{V}^n}\sum_{\mathbf{q}_1,\ldots,\mathbf{q}_n}\Gamma^{(n)}(\mathbf{q}_1,\ldots,\mathbf{q}_n)V(\mathbf{q}_1)\ldots V(q_n)$$
$$\times S^{\text{i}}(\mathbf{q}_1,\ldots,\mathbf{q}_n)\Delta(\mathbf{q}_1+\mathbf{q}_n). \quad (13)$$

Here, $S^{\text{i}}(\mathbf{q}_1,\ldots,\mathbf{q}_n)$ is the $n$-particle structure factor of the proton subsystem, which depends only on the proton coordinates and formally accurately takes into account the proton–proton interaction; $\Delta(\mathbf{q}_1+\ldots+\mathbf{q}_n)$ is the Kronecker delta; and $\Gamma^{(n)}(\mathbf{q}_1,\ldots,\mathbf{q}_n)$ is an electron $n$-pole [12], which depends only on the coordinates of the electron subsystem and formally accurately takes into account the electron–electron interaction. The last expression is formally accurate; therefore, it cannot be used for concrete calculations. There are a few versions of approximate calculations both for electron multipoles [15–18] and multiparticle structure factors of the proton subsystem [19]. For an electron two-pole, the result, common to all authors, is as follows:

$$\Gamma^{(2)}(\mathbf{q},-\mathbf{q}) = -\frac{1}{2}\frac{\pi(q)}{\varepsilon(q)}. \quad (14)$$

] Here, $\pi(q)$ is a polarization function of the electron gas and $\varepsilon(q)$ is its dielectric permittivity. In the random-phase approximation, when the exchange interaction and the electron correlation in the local-filed approximation are taken into account, we have

$$\varepsilon(q) = 1 + [V(q) + U(q)]\pi_0(q), \quad (15)$$

where

$$U(q) = -\frac{2\pi e^2}{q^2 + \lambda k_{\text{F}}^2}$$

is the potential energy of the exchange interaction and the electron-gas correlations, $\lambda \approx 2$ [20], $\pi_0(q)$ is a polarization function of an ideal electron gas. For an electron three-pole, the results obtained by different authors are essentially different. The calculation of an electron three-pole, performed independently by the present author for the model of an ideal electron gas, yielded the same result as that obtained in [17]. It is this result that we use in the present paper:

$$\Gamma^{(3)}(\mathbf{q}_1,\mathbf{q}_2,\mathbf{q}_3) = \frac{\Lambda_0^{(3)}(\mathbf{q}_1,\mathbf{q}_2,\mathbf{q}_3)}{\varepsilon(\mathbf{q}_1)\varepsilon(\mathbf{q}_2)\varepsilon(\mathbf{q}_3)}, \quad (16)$$

where $\Lambda_0^{(3)}(\mathbf{q}_1,\mathbf{q}_2,\mathbf{q}_3)$ is an electron three-pole of a degenerate ideal electron gas. The above-mentioned approximation for a three-pole corresponds to taking into account the electron–electron interaction in the self-consistent-field approximation, where the electron–electron approximation is taken into consideration only through the shielding of the external field—the field of protons. The second- and third-order terms in the electron–proton approximation are expressed as follows after passing from summation to integration in spherical coordinates:

$$E_2 = N\frac{-1}{4\pi^2}\int_0^\pi \frac{\pi(q)}{\varepsilon(q)}w^2(q)S(q)q^2 dq, \quad (17)$$

$$E_3 = N\frac{1}{4\pi^4}\int_0^\infty dq_1 q_1^2 \int_0^\infty dq_2 q_2^2 F(q_1,q_2), \quad (18)$$

$$F(q_1,q_2) = \frac{2n+1}{2}\int_0^\pi \frac{\Lambda_0^{(3)}(\mathbf{q}_1,\mathbf{q}_2,-\mathbf{q}_1,-\mathbf{q}_2)}{\varepsilon(q_1)\varepsilon(q_2)\varepsilon(|\mathbf{q}_1-\mathbf{q}_2|)}$$
$$\times w(q_1)w(q_2)w(|\mathbf{q}_1+\mathbf{q}_2|)S(\mathbf{q}_1,\mathbf{q}_2,-\mathbf{q}_1,-\mathbf{q}_2)\sin\theta_{12}d\theta_{12}.$$

$$S(\mathbf{q}_1,\mathbf{q}_2,\mathbf{q}_3) = S(\mathbf{q}_1)S(\mathbf{q}_2)S(\mathbf{q}_3). \quad (19)$$

Since the electron–proton interaction is known exactly, the main approximation that we use when calculating the third-order term in the electron–proton interaction is the geometrical approximation for a three-particle structure factor [19, 21, 22]:

$$S(\mathbf{q}_1,\mathbf{q}_2,\mathbf{q}_3) = S(\mathbf{q}_1)S(\mathbf{q}_2)S(\mathbf{q}_3). \quad (20)$$

Thus, the ground-state energy of electron gas in metallic hydrogen can be expressed as

$$E = E_0 + \sum_{n=2}^\infty E_n, \quad (21)$$

$$E_0 = E_{0\text{e}} + E_{\text{HF}} + E_{\text{cor}}. \quad (22)$$

Figure 1 shows that, as the density increases, the role of the band structure energy (the second- and third-order terms in the electron–proton interaction) decreases, thus facilitating the convergence of the perturbation-theory series with respect to this interaction. In addition, throughout the density and temperature ranges considered, the third-order term in the electron–proton interaction is much smaller than the second-order term. Another noteworthy fact is that, for densities higher than the densities of transition to the metallic state (0.3 mol/cm³), the internal energy becomes positive

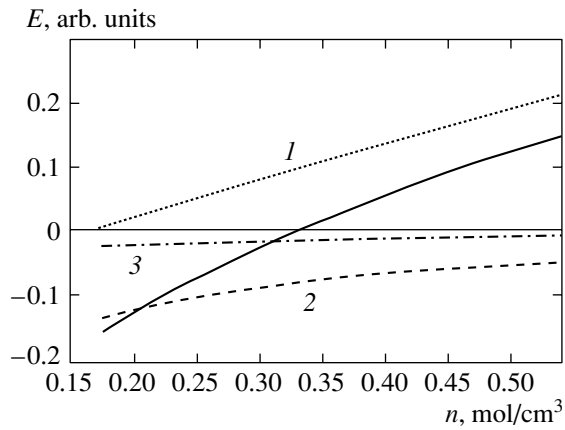

**Fig. 1.** Ground-state energy of electron gas at temperature of 9000 K. $E_0$ (1), $E_2$ (2), and $E_3$ (3) are the contributions of the zero-, second-, and third-order terms in the electron–proton interaction to the energy. The solid curve corresponds to $E_0$ (1) + $E_2$ (2) + $E_3$ (3).

and approaches, as density increases, the internal energy of an ideal gas.

## 4. FREE ENERGY AND PRESSURE

By definition, the free energy is

$$F = E - TS, \qquad (23)$$

where $S$ is the entropy of a system. The entropy can be represented as a sum of electron and proton components. However, for a degenerate electron gas, we can neglect the electron component compared with the proton contribution to the entropy; in the hard-spheres approximation [14, 23], the latter contribution can be expressed as

$$S = S_i = S_{hs} = S_{0i} + S_i(\eta), \qquad (24)$$

where

$$S_{0i} = Nk_B \ln\left[\frac{e}{n}\left(\frac{eMk_BT}{2\pi\hbar^2}\right)^{3/2}\right] \qquad (25)$$

is the entropy of an ideal proton gas ($M$ is the proton mass, and $n$ is the density of protons) and

$$S_i(\eta) = Nk_B \frac{3\eta^2 - 4\eta}{(1-\eta)^2} \qquad (26)$$

is a contribution due to the proton–proton interaction.

The theory proposed contains, at first glance, the only undetermined parameter, the diameter of hard spheres. The knowledge of this parameter as a function of density and temperature is necessary for determining the corresponding functions of thermodynamic potentials. To find the latter parameters, we used an effective pair proton–proton interaction. Using the dependence of this interaction on the proton–proton distance, one can determine the diameter of hard spheres for arbitrary temperatures and densities [10]. The only approximation used when deriving this dependence is the above-discussed random-phase approximation for the electron subsystem with regard to the exchange interaction and electron correlation in the local-field approximation.

The diameter of hard spheres, i.e., the minimal distance to which protons can approach each other at a given temperature, is determined from the equality of the kinetic and potential energies of protons as they approach each other:

$$V_{\text{eff}}(\sigma) = \frac{3}{2}k_BT. \qquad (27)$$

Calculations show that, at the density corresponding to the transition density of hydrogen to the metallic state (0.3 mol/cm³), the depth of the potential well is only a few hundred degrees. At higher densities, the potential well virtually disappears. Thus, at high temperatures, only repulsion of protons is significant in metallic hydrogen. This fact makes impossible the existence of metallic hydrogen at high temperatures when the external pressure is removed.

## 5. DISCUSSION OF THE RESULTS

According to the discoverers of metallic hydrogen [2], at a temperature of 3000 k and density of 0.3 mol/cm³, the pressure amounts to 1.4 Mbar. Our calculations yield a pressure of 1.38 Mbar under the same conditions. In our view, such close values point not so much to the adequacy of the theory to experimental conditions as to the similarity between the calculation methods for pressure and the simplifying assumptions made. An answer to the question of the applicability of, say, the nearly free-electron model must be sought within the theory itself. A dimensionless parameter that characterizes the applicability of the nearly free-electron model is $\hbar/\varepsilon_F\tau$, where $\tau$ is the lifetime of an electron in the state with a given wave vector. This lifetime is close to the relaxation time for the electric conductivity and thermal conductivity of metals. The nearly free-electron model can be applied in the case when this parameter is less than unity. Figure 2 represents the above-mentioned parameter as a function of density and temperature. Figure 2 also shows that, at small densities and high temperatures, the dimensionless parameter is close to unity; i.e., the nearly free-electron model is inapplicable in this case. For high densities and low temperatures, the parameter is much less than unity, and the nearly free-electron model works well. In particular, the nearly free-electron model works quite well for a density of $n = 0.3$ mol/cm³, at which metallic hydrogen was first obtained.

Another important moment of the theory is the essential role played by the electron–proton interaction in the formation of not only kinetic but also thermodynamic properties of metallic hydrogen. This is illustrated by the following results of numerical calcula-tions. When the electron–proton interaction is not taken into account, at a temperature of 3000 K and density of 0.3 mol/cm³, the pressure has an unphysical value of –0.27 Mbar. If we take into account the electron–proton interaction in the second-order of perturbation theory, we obtain a pressure of 1.13 Mbar. If we additionally take into account the third-order term, we obtain a pressure of 1.38 Mbar.

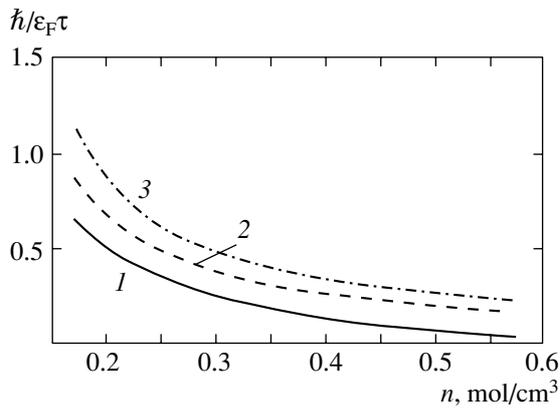

**Fig. 2.** The dimensionless parameter $\hbar/\varepsilon_F\tau$ as a function of density at various temperatures; (*1*) $T = 3000$ K, (*2*) $T = 9000$ K, and (*3*) $T = 18\,000$ K.

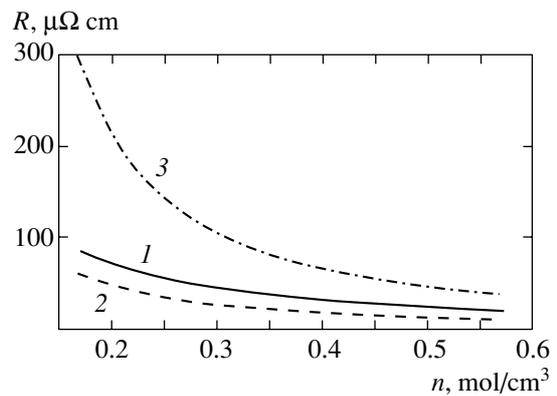

**Fig. 3.** Electric resistivity as a function of density at temperature of 9000 K; $R_2$ (*1*) and $R_3$ (*2*) are the second- and third-order terms and $R$ (*3*) is the result of approximate summation of the perturbation theory series.

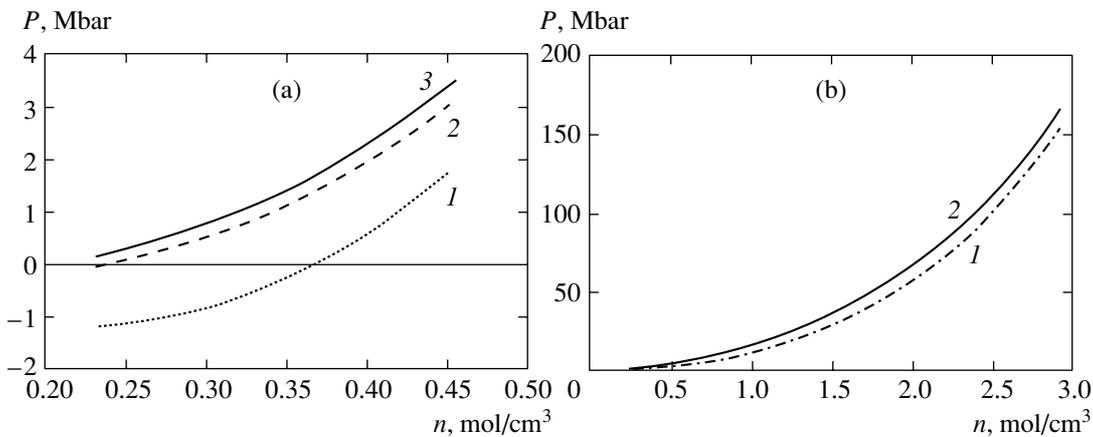

**Fig. 4.** The pressure of metallic hydrogen as a function of density at temperatures of (a) $T = 3000$ K and (b) 20 000 K. $P_0$ (*1*) is the pressure of hydrogen without taking the electron–proton interaction into account, $P_2$ is the pressure due to the contribution of the second-order term in the electron–proton interaction, and $P_3$ is the pressure due to the contribution of the third-order term in the electron–proton interaction. Curve 2 corresponds to the sum $P_0$ and $P_2$, and curve 3, to the sum $P_0 + P_2 + P_3$.

The perturbation theory series for energy converges much better than, say, for the electric resistivity [10]. In the latter case, the perturbation theory series, which begins with the second-order terms in the electron–proton interaction, converges rather slowly (see Fig. 3). Figure 3 also shows that the third-order term is less than the second-order term only by a few tens of percent throughout the density range. The convergence of the perturbation theory series for electric resistivity increases as the density increases.

As for the pressure, it grows both with temperature and density. The corresponding functions are monotone and nonlinear. This is clearly illustrated by Fig. 4.

Figures 4a and 4b also show that the role of the electron–proton interaction is rather significant at relatively low temperatures and densities, say, at temperature of 3000 K. At temperature of 20000 K, the role of the contribution of the third-order term in the electron–proton interaction to pressure is negligible. At this temperature, the role of the contribution of the second-order term also substantially decreases. In this case, the equation of state of metallic hydrogen differs slightly from the equation of state of the mixture of electronic and ionic ideal gases.

One can easily determine the domain of existence of a liquid metallic phase in the hard-spheres model. Transition of the system to a crystalline state corresponds to the situation when the packing fraction of hard spheres approaches the maximum possible value [23]. At temperature of 3000 K, the limit value of the density of the liquid phase is approximately equal to 0.5 mol/cm$^3$, while, at temperature of 9000 K, it equals 1.6 mol/cm$^3$, etc. Note that, under surmised values of density and temperature in the core of Jupiter, liquid metallic hydrogen is rather far from the transition point to the solid state.

The reliability of the results obtained at temperatures far above 3000 K is much higher than that at the temperature at which metallic hydrogen was produced under terrestrial conditions. The reason is that, at a density of 0.3 mol/cm$^3$, there is a 3000-K band-gap in the electron energy spectrum of metallic hydrogen. For high temperatures, one should take into account temperature corrections to multipoles and to the permittivity of the electron gas. It is obvious that, even in the temperature range considered, the dimensionless parameter $k_B T/\varepsilon_F$, which characterizes the degeneracy of the electron gas, is rather large, and its contribution may appreciably change the results obtained.

Returning to the value of pressure at which metallic hydrogen was discovered, we note the following. We have calculated the electric resistivity and pressure within the same model and in identical approximations. The experimental values of the electric resistivity are much higher than those obtained by approximate summation of the perturbation theory series for the electric resistivity. Under these conditions, the theoretical values of pressure obtained by us cannot be close to the experimental values. Such uncertainty is not associated with the model of metallic hydrogen used. As pointed out above, the applicability conditions of the model are quite well satisfied. One reason is the neglect of the existence of the band-gap in the electronic energy spectrum, which is only essential at relatively low temperatures and densities. Another reason is that it is unknown what part of hydrogen is in atomic state. This reason may also be essential only at relatively low temperatures and densities.

Thus, the theory proposed in this paper pretends to a quantitative description of both equilibrium and nonequilibrium properties of metallic hydrogen only in the range of high temperatures and high densities. If temperature is equal to 3000 K, the density must be much higher than 0.3 mol/cm$^3$, and, if the density is equal to 0.3 mol/cm$^3$, the temperature must be much higher than 3000 K.

SPELL: 1. unphysical